\documentclass[useAMS,usenatbib]{mn2e}

\usepackage{graphicx,amssymb,amsmath,color}

\title[MW satellites and CDM-radiation interactions]
{Using the Milky Way satellites to study interactions between cold dark matter and radiation}
\author[C. B\oe hm et al.]
{C.~B\oe hm,$^{1,2}$\thanks{E-mail: c.m.boehm@durham.ac.uk}
J.~A.~Schewtschenko,$^{1,3}$
R.~J.~Wilkinson,$^1$
C.~M.~Baugh,$^3$
S.~Pascoli$^1$\thanks{Also visiting Instituto de F\'{\i}sica Te\'orica, IFT-UAM/CSIC, Universidad Aut\'onoma de Madrid, Cantoblanco, 28049, Madrid, Spain.}\vspace*{6pt}\\
$^1$Institute for Particle Physics Phenomenology, Durham University, Durham DH1 3LE, UK\\
$^2$LAPTH, U. de Savoie, CNRS,  BP 110, 74941 Annecy-Le-Vieux, France\\
$^3$Institute for Computational Cosmology, Durham University, Durham DH1 3LE, UK}

\date{\today}

\pubyear{2014}

\begin{document}

\label{firstpage}

\maketitle


\begin{abstract}
The cold dark matter (CDM) model faces persistent challenges on small scales. In particular, taken at face value, the model significantly overestimates the number of satellite galaxies around the Milky Way. Attempts to solve this problem remain open to debate and have even led some to abandon CDM altogether. However, current simulations are limited by the assumption that dark matter feels only gravity. Here, we show that including interactions between CDM and radiation (photons or neutrinos) leads to a dramatic reduction in the number of satellite galaxies, alleviating the Milky Way satellite problem and indicating that physics beyond gravity may be essential to make accurate predictions of structure formation on small scales. The methodology introduced here gives constraints on dark matter interactions that are significantly improved over those from the cosmic microwave background.
\end{abstract}


\section{Introduction}
\label{sec:intro}

$N$-body simulations of `cold' dark matter (CDM), consisting of weakly-interacting particles with a low velocity dispersion and therefore, negligible free-streaming, agree remarkably well with observations of the Universe on the largest scales~\citep{Davis:1985rj}. However, as the resolution of the simulations improved, significant discrepancies emerged on small scales. For example, dark matter (DM) halo profiles for dwarf galaxies are less cuspy than predicted by CDM~(\citealt{Dubinski:1991bm}, although this is still under debate, see~\citealt{Frenk:2012ph}) and large CDM haloes do not form as many stars as expected (the `too big to fail problem';~\citealt{BoylanKolchin:2011de}).

Here, we address the so-called `Milky Way satellite problem'~\citep{Klypin:1999uc,moore_dark_1999}, which describes the disagreement between the number of `satellite' galaxies in orbit around the Milky Way (MW) and the much larger abundance of DM subhaloes predicted by the CDM model. Whilst the observational data have been revised significantly since this problem was first discussed and their completeness is still under discussion (e.g.~\citealt{Tollerud:2008ze}), a clear discrepancy remains compared with the number of DM subhaloes around the MW.

Several astrophysical processes have also been invoked to solve this problem. Star formation could have been suppressed due to the effects of supernova feedback, photoionization and reionization~\citep{Bullock:2000wn,Benson:2001au} and tidal stripping may have dramatically reduced the size of substructures or disrupted a fraction of them~\citep{Kravtsov:2004cm}. Alternatively, since the DM halo mass has a significant impact on the expected number of satellites, one may argue that the severity of the problem depends upon the choice of MW halo mass, which remains difficult to determine~\citep{2012MNRAS.424.2715W,Cautun:2014dda}.

A more drastic solution is to abandon CDM and instead consider `warm' dark matter (WDM). In this scenario, one allows a small (but non-negligible) amount of free-streaming, which greatly reduces the expected number of satellites with respect to CDM~\citep{Lovell:2013ola}. Given that the free-streaming scale of a DM particle is typically governed by its mass and velocity distribution, the proposed WDM models require very light ($\sim$ keV) particles. However, recent work suggests that such light candidates cannot simultaneously solve the small-scale problems of CDM and satisfy the particle mass constraints from the Lyman $\alpha$ forest and other observations~\citep{Schneider:2013wwa,Viel:2013fqw}.

Here, we explore an alternative route that allows us to reduce the MW satellite population without having to discard CDM. In standard $N$-body simulations, DM is represented as a collisionless fluid that responds only to gravity. However, it is entirely plausible (and indeed expected) that DM interacts through other forces, with various components of the Universe\footnote{We do not study self-interactions as these have been discussed already in~\citet{rocha_cosmological_2012}.}. Such interactions have been shown to suppress small-scale density fluctuations~\citep{boehm_constraining_2001,boehm_interacting_2001,Boehm:2003xr,Boehm:2004th,chen_cosmic_2002,2012PhRvL.109w1301V,dvorkin_constraining_2013} but the implications for the satellite galaxy abundance have not been studied using numerical simulations.

Here, we simulate the formation of large-scale structure in a Universe where DM interacts with photons or neutrinos, to determine whether such a coupling can address the MW satellite problem. We focus on radiation as this dominates the energy density at early times and should therefore lead to the largest effect on DM primordial fluctuations. For the sake of illustration, we will study specifically a DM--photon coupling (hereafter referred to as $\gamma$CDM) but very similar effects are expected in the case of a DM--neutrino coupling ($\nu$CDM). We will use these results to extract constraints on the DM--photon scattering cross section. The other small-scale problems of CDM will be addressed in forthcoming work.

The Letter is organized as follows. In Section~\ref{sec:background}, we discuss the theoretical framework of interacting DM models. In Section~\ref{sec:simulations}, we provide details regarding the setup of our simulations. In Section~\ref{sec:results}, we present the results of our simulations and in particular, the effect on the satellite galaxy abundance. Conclusions are provided in Section~\ref{sec:conc}.

\section{Interacting DM}
\label{sec:background}

Regardless of the particle physics model, DM interactions beyond gravity result in an additional collision term in the corresponding Boltzmann equations. For $\gamma$CDM, the modified Boltzmann equations read
\begin{eqnarray}
\dot \theta_{\rm DM} &=& k^2 \psi - \mathcal{H} \theta_{\rm DM} - S^{-1} \dot \mu (\theta_{\rm DM} - \theta_\gamma) ~, \\
\dot \theta_{\gamma} &=& k^2 \psi + k^2 \left( \frac 1 4 \delta_\gamma - \sigma_\gamma \right) -\dot \mu (\theta_\gamma-\theta_{\rm DM})~,
\end{eqnarray}
where $\theta$ is the velocity dispersion, $k$ is the wavenumber, $\psi$ is the (DM-dominated) gravitational potential, $\mathcal{H}$ is the expansion rate of the Universe, $\delta$ is the density contrast and $\sigma$ is the anisotropic stress potential\footnote{We use the Newtonian gauge, where the Thomson scattering terms are omitted for brevity.}~\citep{wilkinson_using_2013}.

The new interaction rate, $\dot \mu$, is the product of the scattering cross section, $\sigma_{\rm{DM}-\gamma}$, and the DM number density, while the DM--photon ratio, $S$, ensures energy conservation. For simplicity, we take $\sigma_{\rm{DM}-\gamma}$ to be constant (however, a temperature-dependent cross section has a similar impact) and assume that the interacting DM species accounts for the entire observed relic abundance.

This formalism provides an accurate estimate of the collisional damping scale associated with DM interactions in the linear regime (when the density fluctuations are small). However, one can understand the underlying physics by considering both the DM and radiation as interacting imperfect fluids~\citep{boehm_constraining_2001, Boehm:2004th} leading to a damping scale
\begin{eqnarray}
l^2_{{\rm cd},\gamma} \sim \int_0^{t_{{\rm dec}({\rm DM}-\gamma)}}  \frac{\rho_{\gamma}~c^2}{\rho~\Gamma_{\gamma}~a^2}~{\rm d}t~,
\label{eq:l2cd}
\end{eqnarray}
where $\rho_{\gamma}$ is the photon energy density, $\rho$ is total energy density, $\Gamma_{\gamma}$ is the total interaction rate of the photons (including all species in thermal equilibrium with them) and $a$ is the cosmological scale factor.

Eq.~\eqref{eq:l2cd} illustrates why interactions with radiation can lead to the suppression of small-scale power needed to  reduce the MW satellite population. In the early Universe, photons and neutrinos were ultrarelativistic and constituted the bulk of the energy density. Hence, the numerator in Eq.~\eqref{eq:l2cd} is large and fluctuations can be erased on the scale of small galaxies, depending on the strength of the interaction.

\begin{figure}
\begin{centering}
\includegraphics[width=8.8cm, trim = 3cm 2.8cm 0.3cm 2cm]{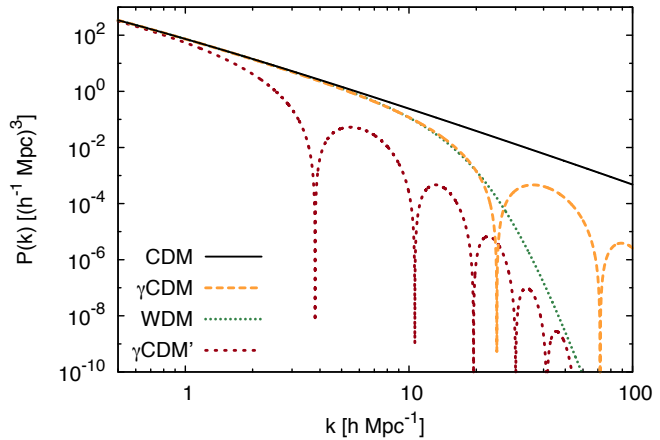}
\caption{The linear matter power spectra for CDM, $\gamma$CDM with $\sigma_{\rm{DM}-\gamma} = 2 \times 10^{-9}~\sigma_{\rm Th}\left(m_{\rm{DM}}/\rm{GeV}\right)$, WDM with $m_{\rm DM} = 1.24~\mathrm{keV}$ and $\gamma$CDM' with $\sigma_{\rm{DM}-\gamma} = 10^{-7}~\sigma_{\rm Th}\left(m_{\rm{DM}}/\rm{GeV}\right)$. We take $\sigma_{\rm{DM}-\gamma}$ to be constant and use the best-fitting cosmological parameters from {\it Planck}~\citep{Ade:2013zuv}.}
\label{fig:P_k}
\end{centering}
\end{figure}

The consequences of DM interactions with radiation have been computed in the linear regime~\citep{boehm_interacting_2001,Sigurdson:2004zp,Wilkinson:2014ksa}. The $\gamma$CDM matter power spectrum is damped relative to that of CDM beyond a scale that depends on the interaction cross section (see Fig.~\ref{fig:P_k}). This is similar to the damping seen in WDM, except that in this case, instead of an exponential suppression, one obtains a series of oscillations with a power law modulation of their amplitude~\citep{boehm_interacting_2001}. We can compare WDM and $\gamma$CDM by choosing particle masses, $m_{\rm WDM}$, and interaction cross sections, $\sigma_{\rm{DM}-\gamma}$, that produce a damping relative to CDM at a similar wavenumber.

For $\gamma$CDM, the comparison with cosmic microwave background (CMB) data from {\it Planck}~\citep{Ade:2013zuv} gives a constraint on the (constant) elastic scattering cross section of $\sigma_{\rm{DM}-\gamma} \lesssim 10^{-6}~\sigma_{\rm Th}\left(m_{\rm{DM}}/\rm{GeV}\right)$, where $\sigma_{\rm Th}$ is the Thomson cross section and $m_{\rm DM}$ is the DM mass~\citep{wilkinson_using_2013}. However, this linear approach breaks down once the fluctuations become large, preventing one from studying the effects of weak interactions on DM haloes and in particular, on small-scale objects.

\section{Simulations}
\label{sec:simulations}

To study small-scale structures in these models, we begin our $N$-body calculations at a sufficiently early epoch ($z=49$), where the effect of $\gamma$CDM is fully described by linear perturbation theory. The DM--photon interaction rate is negligible for $z < 49$. The initial matter power spectra are obtained from a modified version of the Boltzmann code \textsc{class}~\citep{class_refs}, using the best-fitting cosmological parameters from {\it Planck}~\citep{Ade:2013zuv}. Initial conditions are created using a second-order LPT code. 

To make predictions in the non-linear regime, we run a suite of $N$-body simulations using the code \textsc{gadget-3}~\citep{springel_cosmological_2005}. To provide a suitable dynamical range, we perform simulations in both a large box ($100~{h}^{-1}~{\rm Mpc}$, $512^3$ particles) and a high-resolution small box ($30~{h}^{-1}~{\rm Mpc}$, $1024^3$ particles). A subset of simulations is re-run in a high-resolution large box ($100~{h}^{-1}~{\rm Mpc}$, $1024^3$ particles) to confirm the convergence scale. By comparing the results from different runs, we find that our calculations are reliable for subhaloes with $V_{\rm max} \gtrsim 8~\rm{km}~{\rm{s}}^{-1}$. Gravitational softening is set to 5 per cent of the mean particle separation. For WDM particles with masses larger than $\sim$ keV, the thermal velocities are sufficiently small that one can safely neglect the free-streaming in the non-linear regime without introducing a significant error on the scales of interest~\citep{colin_on_2008}.

To quantify the impact of $\gamma$CDM on MW satellites, one needs to define criteria to select haloes that could host the MW. The most crucial condition is the DM halo mass. Motivated by calculations that attempt to reconstruct the MW mass distribution based on the measured kinematics of the observed satellites and stars~\citep{xue_milky_2008,boylan-kolchin_space_2012-1,piffl_rave_2013}, we consider DM haloes to be MW-like if their mass is in the range $(0.8 - 2.7) \times 10^{12} {\rm M}_\odot$.

The second criterion we apply is based on environment. The MW appears to be located in an unremarkable region away from larger structures such as the Virgo Cluster and the major filaments feeding the Centaurus Cluster~\citep{courtois_cosmography_2013}. We therefore reject candidates with similar-sized haloes within a neighbourhood of 2 Mpc. The resulting sample of MW-like haloes is then divided into several subsets based on their virial halo mass. Haloes are identified using a friends-of-friends group finder~\citep{Davis:1985rj} with a linking length of 0.2 times the mean particle separation and subhaloes are identified using \textsc{subfind}~\citep{Springel:2000yr}.

\section{Results}
\label{sec:results}

\begin{figure*}
\includegraphics[width=11cm]{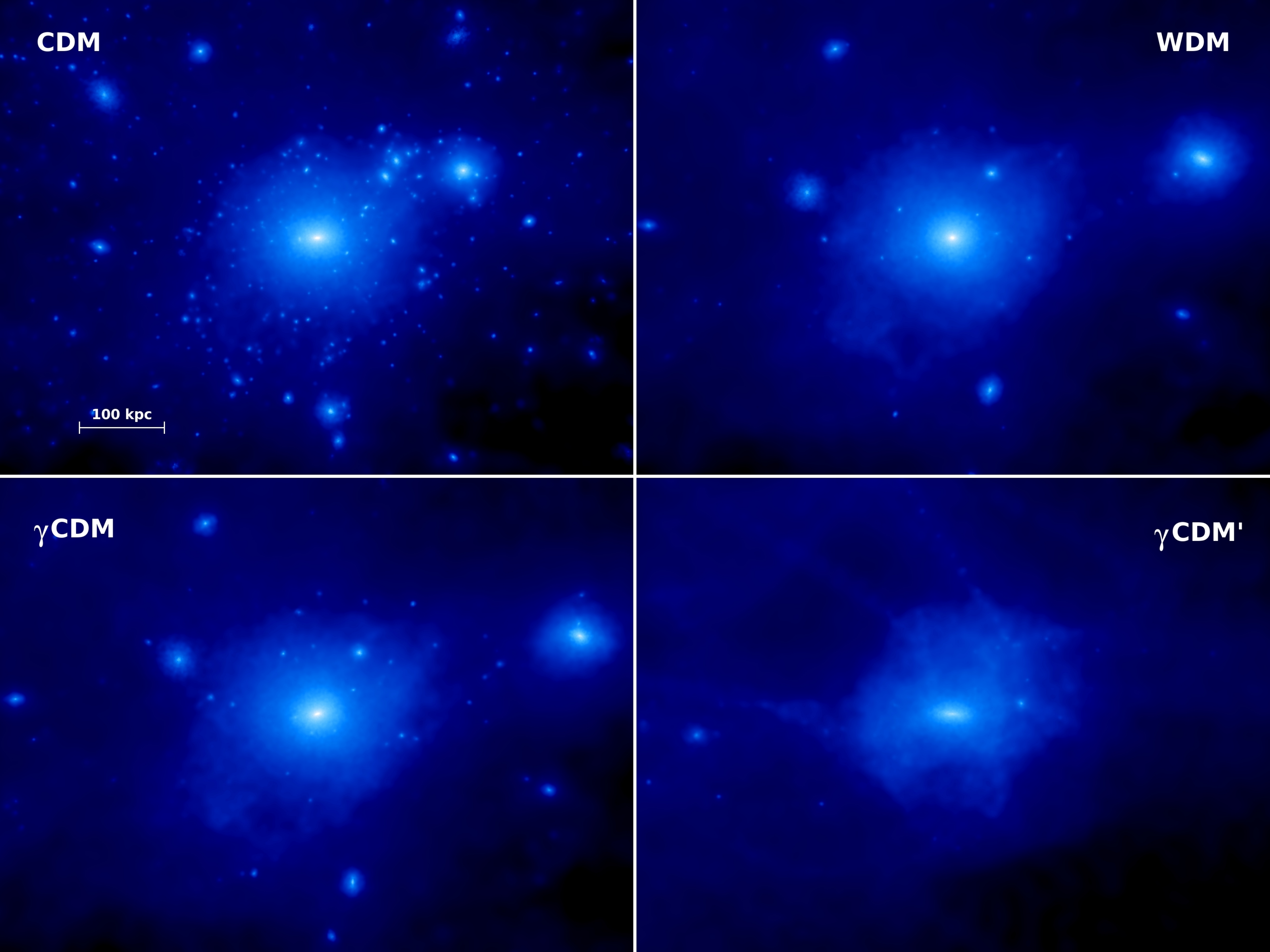}
\vspace{1.5ex}
\caption{The simulated distribution of DM in a MW-like halo. The shading represents the DM density, with brighter colours indicating higher densities. The panels show the halo in simulations of different cosmological models: CDM (top left), $\gamma$CDM with $\sigma_{\rm{DM}-\gamma} = 2 \times 10^{-9}~\sigma_{\rm Th}\left(m_{\rm{DM}}/\rm{GeV}\right)$ (bottom left), the equivalent model of WDM with $m_{\rm DM} = 1.24~\mathrm{keV}$ (top right) and $\gamma$CDM' with $\sigma_{\rm{DM}-\gamma} = 10^{-7}~\sigma_{\rm Th}\left(m_{\rm{DM}}/\rm{GeV}\right)$ (bottom right). The large number of subhaloes observed in the top-left panel illustrates the MW satellite problem. By replacing CDM with WDM (top right), the number of subhaloes is reduced dramatically. A similar paucity of subhaloes is seen in the bottom-right panel, in which the DM--photon interaction strength is just allowed by CMB constraints~\citep{wilkinson_using_2013}. This model underestimates the number of MW satellites. The model in the bottom-left panel has an interaction strength that is 1000 times smaller than the CMB limit, in which the number of subhaloes is a much better match to the observed number of satellites.}
\vspace{2ex}
\label{fig:sims}
\end{figure*}

Fig.~\ref{fig:sims} shows the simulated distribution of DM in a MW-sized DM halo. For CDM (top-left panel), there is a large abundance of subhaloes within the DM halo, which illustrates the MW satellite problem. The bottom-left panel shows the same halo in a simulation of $\gamma$CDM, in which the interaction cross section is $\sigma_{\rm{DM}-\gamma} = 2 \times 10^{-9}~\sigma_{\rm Th}\left(m_{\rm{DM}}/\rm{GeV}\right)$. Such a cross section should satisfy the constraints from the Lyman $\alpha$ forest~\citep{Viel:2013fqw}. The subhalo population is significantly smaller for this model compared to CDM. However, the suppression of subhaloes in $\gamma$CDM is too strong if we consider $\sigma_{\rm{DM}-\gamma} = 10^{-7}~\sigma_{\rm Th}\left(m_{\rm{DM}}/\rm{GeV}\right)$ (bottom-right panel), which just satisfies the limit from the CMB~\citep{wilkinson_using_2013}. Therefore, by adjusting the magnitude of the scattering cross section, not only is there scope to address the MW satellite problem, but we can also place a more stringent constraint on the $\gamma$CDM interaction strength.

For the model of $\gamma$CDM with $\sigma_{\rm{DM}-\gamma} = 2 \times 10^{-9}~\sigma_{\rm Th}\left(m_{\rm{DM}}/\rm{GeV}\right)$, the distribution of density fluctuations in the linear regime is similar to that of a WDM particle with a mass of 1.24~keV (top-right panel). However, the suppression of small-scale power in $\gamma$CDM is less extreme than in generic WDM models due to the presence of oscillations in the power spectrum (see Fig.~\ref{fig:P_k}), which offers a way to distinguish these two scenarios (and possibly provide a solution to the `too big to fail' problem).

\begin{figure*}
\includegraphics[width=17cm]{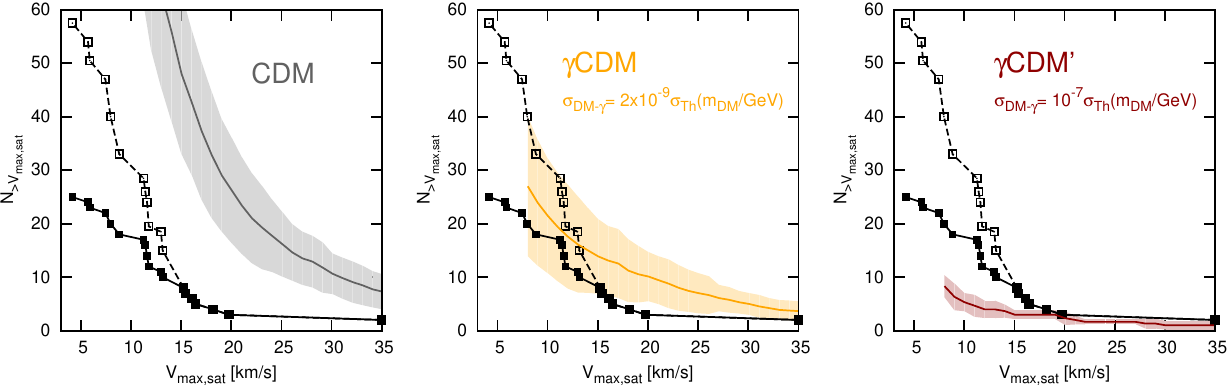}
\vspace{1ex}
\caption{The number of satellite galaxies in a MW-like DM halo as a function of their maximal circular velocity: CDM (left), $\gamma$CDM with $\sigma_{\rm{DM}-\gamma} = 2 \times 10^{-9}~\sigma_{\rm Th}\left(m_{\rm{DM}}/\rm{GeV}\right)$ (middle) and $\gamma$CDM' with $\sigma_{\rm{DM}-\gamma} = 10^{-7}~\sigma_{\rm Th}\left(m_{\rm{DM}}/\rm{GeV}\right)$ (right). The lines and shading show the mean cumulative number counts of MW satellites for a simulated DM halo in the mass bin $(2.3-2.7) \times 10^{12} {\rm M}_\odot$ and the 1$\sigma$ uncertainty. Also plotted are the observational results~(\citealt{Willman:2009dv}, solid black lines), which are then corrected for the completeness of the Sloan Digital Sky Survey coverage (dashed lines). The maximal circular velocity, $V_{\rm max}$, is selected as a measure for the mass and is determined directly from the simulations (it is derived from the observed stellar line-of-sight velocity dispersions using the assumption that $V_{\rm max} = \sqrt{3} \sigma_{\star}$;~\citealt{Klypin:1999uc}). The number of selected MW-like haloes are 11, 13 and 3 for CDM, $\gamma$CDM and $\gamma$CDM', respectively (the reduced scatter for $\gamma$CDM' is simply a result of the small-number statistics in this extreme model).}
\label{fig:counts}
\end{figure*}

For more quantitative estimates, the cumulative number counts of MW satellites are plotted in Fig.~\ref{fig:counts} as a function of their maximal circular velocity. The simulation results are obtained by averaging over the haloes that satisfy the selection criteria outlined above. The left-hand panel shows predictions for the CDM model, with no interactions, in which the predicted number of subhaloes of a given maximum circular velocity greatly exceeds the observed number of MW satellites. The middle panel shows the results for $\sigma_{\rm{DM}-\gamma} = 2 \times 10^{-9}~\sigma_{\rm Th}\left(m_{\rm{DM}}/\rm{GeV}\right)$, where there is a good match to the observed number of satellites. Thus, we see that $\gamma$CDM with a small cross section can alleviate the MW satellite problem. Finally, the right-hand panel of Fig.~\ref{fig:counts} shows the model of $\gamma$CDM with $\sigma_{\rm{DM}-\gamma} = 10^{-7}~\sigma_{\rm Th} \times \left(m_{\rm{DM}}/\rm{GeV}\right)$ and clearly, there are not enough small structures remaining in this case.

\begin{figure}
\begin{centering}
\includegraphics[width=7.3cm, trim = -0.4cm -0.4cm 0cm 0cm]{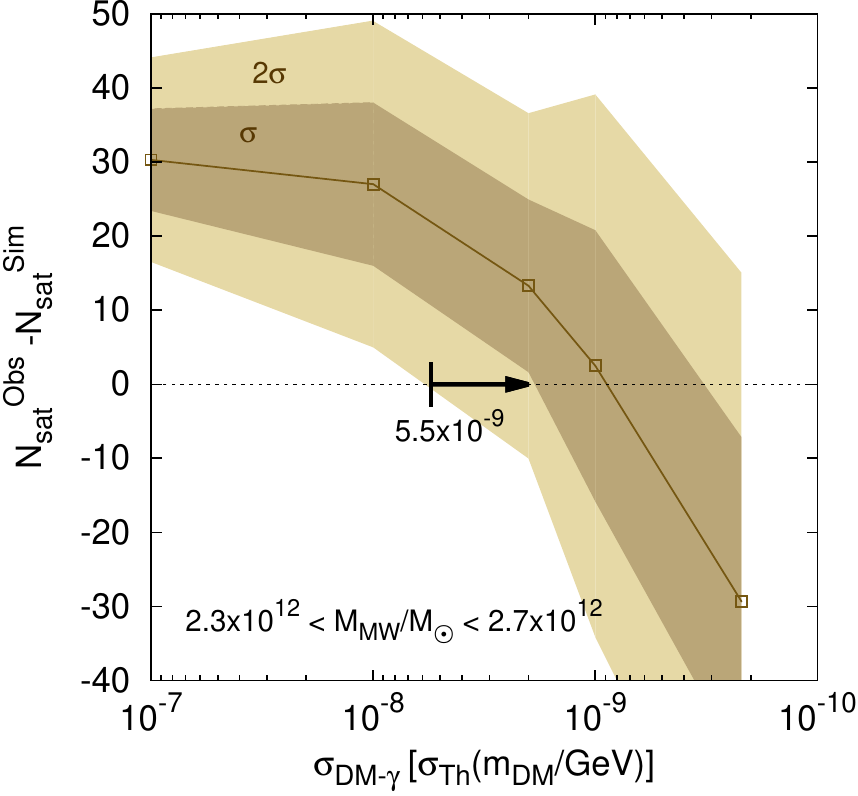}
\includegraphics[width=6.8cm, trim = 0cm 0cm 0cm 0cm]{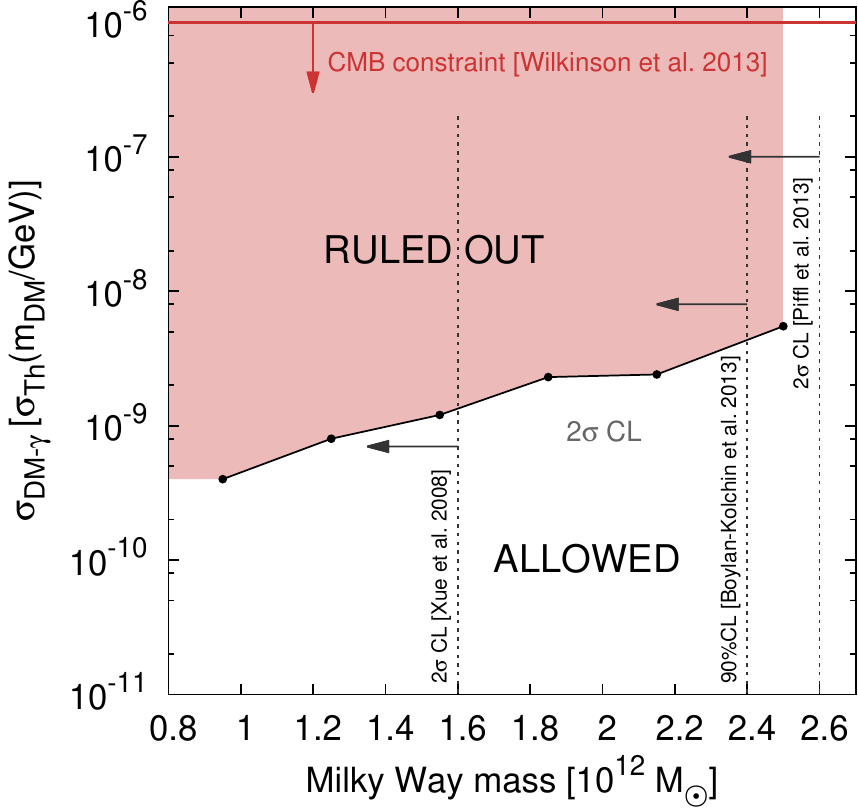}
\caption{Constraints on the $\gamma$CDM cross section. Top panel: the overabundance of satellites versus the cross section for the MW halo mass bin $(2.3-2.7) \times 10^{12} {\rm M}_\odot$, where the shaded bands represent the 1$\sigma$ and 2$\sigma$ uncertainties. Bottom panel: constraints on the cross section are plotted with respect to the MW halo mass. The most recent CMB constraint~\citep{wilkinson_using_2013} and selected upper mass bounds for the MW halo are shown for comparison.}
\label{fig:constraints}
\end{centering}
\vspace{-2ex}
\end{figure}

We can therefore constrain the interaction cross section by comparing the observed and predicted numbers of substructures. The uncertainties in the simulation results are derived from the spread in the sample set (for each host halo mass bin). A given model is ruled out if the number of predicted subhaloes is smaller than the observed number, within the combined uncertainties of these observables (see Fig.~\ref{fig:constraints}, top panel). From this, we conclude that the cross section cannot exceed $\sigma_{\rm{DM}-\gamma} = 5.5 \times 10^{-9}~\sigma_{\rm Th}~(m_{\rm DM}/{\rm GeV})$ (at $2 \sigma$ CL). Note that using the highest mass bin $(2.3-2.7) \times 10^{12} {\rm M}_\odot$ provides us with the most conservative limit. Smaller MW-like halo masses (see Fig.~\ref{fig:constraints}, bottom panel) result in stronger upper bounds on the cross section as these haloes host fewer satellites.

\section{Conclusions}
\label{sec:conc}

We have shown that studying the formation of cosmic structure, particularly on small scales, provides us with a powerful new tool to test the weakly-interacting nature of DM. By performing the first accurate cosmological simulations of DM interactions with radiation (in this case, photons), we find a new means to reduce the population of MW subhaloes, without the need to abandon CDM. The resulting constraints on the interaction strength between DM and photons are orders of magnitude stronger than is possible from linear perturbation theory considerations. Similar results are expected in the case of DM--neutrino interactions.

It should be noted that the observed value of $V_{\rm max}$ may be underestimated by our approach of directly calculating it from the stellar velocity dispersion~\citep{bullock_notes_2010}. Combined with an expected increase in the number of satellites from additional completeness corrections, this would lead to even stricter constraints on the interaction cross section. A future paper will present the non-linear structure formation for such models in greater depth to examine whether one can solve the other small-scale problems of CDM (Schewtschenko et al. 2014).

Recent simulations with DM and baryons have shown that baryonic physics can alter the appearance of the subhalo mass function~\citep{Sawala:2014baa}. A definitive calculation would include the full impact of these effects, in particular, supernovae feedback and photoionization heating of the interstellar medium, but this is deferred to a future paper.

\section*{Acknowledgements}

We thank T. Sawala for comments and discussions, and V. Springel and J. Helly for providing access to the codes used for this work. JAS is supported by a Durham University Alumnus Scholarship and RJW is supported by the STFC Quota grant ST/K501979/1. This work was supported by the STFC (grant numbers ST/F001166/1 and ST/G000905/1) and the European Union FP7 ITN INVISIBLES (Marie Curie Actions, PITN-GA-2011-289442). It made use of the DiRAC Data Centric system at Durham University, operated by the ICC on behalf of the STFC DiRAC HPC Facility (www.dirac.ac.uk). This equipment was funded by BIS National E-infrastructure capital grant ST/K00042X/1, STFC capital grant ST/H008519/1, STFC DiRAC Operations grant ST/K003267/1 and Durham University. DiRAC is part of the National E-Infrastructure. SP also thanks the Spanish MINECO (Centro de excelencia Severo Ochoa Program) under grant SEV-2012-0249.

\label{lastpage}

\bibliographystyle{mn2e}
\bibliography{scdm.bib}

\begin{thebibliography}{}

\bibitem[\protect\citeauthoryear{Ade et~al.,}{Ade  et~al.}{2013}]{Ade:2013zuv}
Ade P.,  et~al., 2013

\bibitem[\protect\citeauthoryear{Benson et~al.,}{Benson
  et~al.}{2002}]{Benson:2001au}
Benson A.~J.,  et~al., 2002, MNRAS, 333, 156

\bibitem[\protect\citeauthoryear{Boehm, Fayet \& Schaeffer}{Boehm
  et~al.}{2001}]{boehm_constraining_2001}
Boehm C.,  Fayet P.,    Schaeffer R.,  2001, Phys. Lett. B, 518, 8

\bibitem[\protect\citeauthoryear{Boehm, Mathis, Devriendt \& Silk}{Boehm
  et~al.}{2005}]{Boehm:2003xr}
Boehm C.,  Mathis H.,  Devriendt J.,    Silk J.,  2005, MNRAS, 360, 282

\bibitem[\protect\citeauthoryear{Boehm, Riazuelo, Hansen \& Schaeffer}{Boehm
  et~al.}{2002}]{boehm_interacting_2001}
Boehm C.,  Riazuelo A.,  Hansen S.~H.,    Schaeffer R.,  2002, Phys. Rev. D,
  66, 083505

\bibitem[\protect\citeauthoryear{Boehm \& Schaeffer}{Boehm \&
  Schaeffer}{2005}]{Boehm:2004th}
Boehm C.,  Schaeffer R.,  2005, A{\&}A, 438, 419

\bibitem[\protect\citeauthoryear{Boylan-Kolchin, Bullock \&
  Kaplinghat}{Boylan-Kolchin et~al.}{2011}]{BoylanKolchin:2011de}
Boylan-Kolchin M.,  Bullock J.~S.,    Kaplinghat M.,  2011, MNRAS, 415, L40

\bibitem[\protect\citeauthoryear{Boylan-Kolchin et~al.,}{Boylan-Kolchin
  et~al.}{2013}]{boylan-kolchin_space_2012-1}
Boylan-Kolchin M.,  et~al., 2013, ApJ, 768, 140

\bibitem[\protect\citeauthoryear{Bullock}{Bullock}{2010}]{bullock_notes_2010}
Bullock J.~S.,  2010

\bibitem[\protect\citeauthoryear{Bullock, Kravtsov \& Weinberg}{Bullock
  et~al.}{2000}]{Bullock:2000wn}
Bullock J.~S.,  Kravtsov A.~V.,    Weinberg D.~H.,  2000, ApJ, 539, 517

\bibitem[\protect\citeauthoryear{Cautun, Frenk, van~de Weygaert, Hellwing \&
  Jones}{Cautun et~al.}{2014}]{Cautun:2014dda}
Cautun M.,  Frenk C.~S.,  van~de Weygaert R.,  Hellwing W.~A.,    Jones B.
  J.~T.,  2014

\bibitem[\protect\citeauthoryear{Chen, Hannestad \& Scherrer}{Chen
  et~al.}{2002}]{chen_cosmic_2002}
Chen X.-l.,  Hannestad S.,    Scherrer R.~J.,  2002, Phys. Rev. D, 65, 123515

\bibitem[\protect\citeauthoryear{Colin, Valenzuela \& Avila-Reese}{Colin
  et~al.}{2008}]{colin_on_2008}
Colin P.,  Valenzuela O.,    Avila-Reese V.,  2008, ApJ, 673, 203

\bibitem[\protect\citeauthoryear{Courtois, Pomarede, Tully \&
  Courtois}{Courtois et~al.}{2013}]{courtois_cosmography_2013}
Courtois H.~M.,  Pomarede D.,  Tully R.~B.,    Courtois D.,  2013, AJ, 146, 69

\bibitem[\protect\citeauthoryear{Davis, Efstathiou, Frenk \& White}{Davis
  et~al.}{1985}]{Davis:1985rj}
Davis M.,  Efstathiou G.,  Frenk C.~S.,    White S.~D.,  1985, ApJ, 292, 371

\bibitem[\protect\citeauthoryear{Dubinski \& Carlberg}{Dubinski \&
  Carlberg}{1991}]{Dubinski:1991bm}
Dubinski J.,  Carlberg R.,  1991, ApJ, 378, 496

\bibitem[\protect\citeauthoryear{Dvorkin, Blum \& Kamionkowski}{Dvorkin
  et~al.}{2014}]{dvorkin_constraining_2013}
Dvorkin C.,  Blum K.,    Kamionkowski M.,  2014, Phys. Rev. D, 89, 023519

\bibitem[\protect\citeauthoryear{Frenk \& White}{Frenk \&
  White}{2012}]{Frenk:2012ph}
Frenk C.,  White S.~D.,  2012, Ann. Phys., Lpz., 524, 507

\bibitem[\protect\citeauthoryear{Klypin, Kravtsov, Valenzuela \& Prada}{Klypin
  et~al.}{1999}]{Klypin:1999uc}
Klypin A.~A.,  Kravtsov A.~V.,  Valenzuela O.,    Prada F.,  1999, ApJ, 522, 82

\bibitem[\protect\citeauthoryear{Kravtsov, Gnedin \& Klypin}{Kravtsov
  et~al.}{2004}]{Kravtsov:2004cm}
Kravtsov A.~V.,  Gnedin O.~Y.,    Klypin A.~A.,  2004, ApJ, 609, 482

\bibitem[\protect\citeauthoryear{Lesgourgues}{Lesgourgues}{2011}]{class_refs}
Lesgourgues J.,  2011

\bibitem[\protect\citeauthoryear{Lovell et~al.,}{Lovell
  et~al.}{2014}]{Lovell:2013ola}
Lovell M.~R.,  et~al., 2014, MNRAS, 439, 300

\bibitem[\protect\citeauthoryear{Moore et~al.,}{Moore
  et~al.}{1999}]{moore_dark_1999}
Moore B.,  et~al., 1999, ApJ, 524, L19

\bibitem[\protect\citeauthoryear{Piffl et~al.,}{Piffl
  et~al.}{2014}]{piffl_rave_2013}
Piffl T.,  et~al., 2014, A{\&}A, 562, A91

\bibitem[\protect\citeauthoryear{Rocha et~al.,}{Rocha
  et~al.}{2013}]{rocha_cosmological_2012}
Rocha M.,  et~al., 2013, MNRAS, 430, 81

\bibitem[\protect\citeauthoryear{Sawala et~al.,}{Sawala
  et~al.}{2014}]{Sawala:2014baa}
Sawala T.,  et~al., 2014

\bibitem[\protect\citeauthoryear{Schneider, Anderhalden, Maccio \&
  Diemand}{Schneider et~al.}{2013}]{Schneider:2013wwa}
Schneider A.,  Anderhalden D.,  Maccio A.,    Diemand J.,  2013

\bibitem[\protect\citeauthoryear{Sigurdson, Doran, Kurylov, Caldwell \&
  Kamionkowski}{Sigurdson et~al.}{2004}]{Sigurdson:2004zp}
Sigurdson K.,  Doran M.,  Kurylov A.,  Caldwell R.~R.,    Kamionkowski M.,
  2004, Phys. Rev. D, 70, 083501

\bibitem[\protect\citeauthoryear{Springel}{Springel}{2005}]{springel_cosmological_2005}
Springel V.,  2005, MNRAS, 364, 1105

\bibitem[\protect\citeauthoryear{Springel, Yoshida \& White}{Springel
  et~al.}{2001}]{Springel:2000yr}
Springel V.,  Yoshida N.,    White S.~D.,  2001, New Astron., 6, 79

\bibitem[\protect\citeauthoryear{Tollerud, Bullock, Strigari \&
  Willman}{Tollerud et~al.}{2008}]{Tollerud:2008ze}
Tollerud E.~J.,  Bullock J.~S.,  Strigari L.~E.,    Willman B.,  2008, ApJ,
  688, 277

\bibitem[\protect\citeauthoryear{{van den Aarssen}, {Bringmann} \&
  {Pfrommer}}{{van den Aarssen} et~al.}{2012}]{2012PhRvL.109w1301V}
{van den Aarssen} L.~G.,  {Bringmann} T.,    {Pfrommer} C.,  2012, Physical
  Review Letters, 109, 231301

\bibitem[\protect\citeauthoryear{Viel, Becker, Bolton \& Haehnelt}{Viel
  et~al.}{2013}]{Viel:2013fqw}
Viel M.,  Becker G.~D.,  Bolton J.~S.,    Haehnelt M.~G.,  2013, Phys. Rev. D,
  88, 043502

\bibitem[\protect\citeauthoryear{Wang, Frenk, Navarro \& Gao}{Wang
  et~al.}{2012}]{2012MNRAS.424.2715W}
Wang J.,  Frenk C.~S.,  Navarro J.~F.,    Gao L.,  2012, MNRAS, 424, 2715

\bibitem[\protect\citeauthoryear{Wilkinson, Boehm \& Lesgourgues}{Wilkinson
  et~al.}{2014}]{Wilkinson:2014ksa}
Wilkinson R.~J.,  Boehm C.,    Lesgourgues J.,  2014, J. Cosmol. Astropart.
  Phys., 1405, 011

\bibitem[\protect\citeauthoryear{Wilkinson, Lesgourgues \& Boehm}{Wilkinson
  et~al.}{2014}]{wilkinson_using_2013}
Wilkinson R.~J.,  Lesgourgues J.,    Boehm C.,  2014, J. Cosmol. Astropart.
  Phys., 1404, 026

\bibitem[\protect\citeauthoryear{Willman}{Willman}{2010}]{Willman:2009dv}
Willman B.,  2010, Adv. Astron., 2010, 285454

\bibitem[\protect\citeauthoryear{Xue et~al.,}{Xue
  et~al.}{2008}]{xue_milky_2008}
Xue X.,  et~al., 2008, ApJ, 684, 1143

\end{thebibliography}

\end{document}